# Resampling Residuals: Robust Estimators of Error and Fit for Evolutionary Trees and Phylogenomics

Peter J. Waddell[12] and Ariful Azad[2]

pwaddell@purdue.edu

[1]Department of Biological Sciences, Purdue University, West Lafayette, IN 47906, U.S.A.
[2]Department of Computer Science, Purdue University, West Lafayette, IN 47906, U.S.A

.

Phylogenomics, even more so than traditional phylogenetics, needs to represent the uncertainty in evolutionary trees due to systematic error. Here we illustrate the analysis of genome-scale alignments of yeast, using robust measures of the additivity of the fit of distances to tree when using flexi Weighted Least Squares. A variety of DNA and protein distances are used. We explore the nature of the residuals, standardize them, and then create replicate data sets by resampling these residuals. Under the model, the results are shown to be very similar to the conventional sequence bootstrap. With real data they show up uncertainty in the tree that is either due to underestimating the stochastic error (hence massively overestimating the effective sequence length) and/or systematic error. The methods are extended to the very fast BME criterion with similarly promising results.

"… his ignorance and almost doe-like naivety is keeping his mind receptive to a possible solution." A quotation from Kryten: Red Dwarf VIII-Cassandra

**Keywords**: Resampled Residuals, Sequence Bootstrap, Flexi Weighted Least Squares or fWLS Phylogenetic Trees, Balanced Minimum Evolution BME, Rokas Yeast Genomes, Phylogenomics

# 1  Introduction

Phylogenetics and phylogenomics aim to recover the evolutionary trees by which homologous parts of the genome evolved. Absent factors such as hybridization or horizontal gene transfer, and if the coalescence time is small compared to edge length durations, then a species tree should be a good description of the data. Phylogenetic methods should aim to not simply build a tree, but inform the user of the fit of the data to the tree (hopefully relative to not only other treatments of the data, but also other data sets) and also the reliability of the inferred tree (e.g., Swofford et al. 1996). One consequence of fit is making a realistic assessment of how much confidence to have that an edge (internode or less precisely a "branch") in the reconstructed tree was is in the tree generating the data. For purposes such as divergence time estimation, with its reliance on the weighted tree, a variance covariance matrix of the edge lengths in the tree is similarly important (e.g., Waddell and Penny 1996, Thorne, Kishino and Painter 1998).

In terms of phylogenetic methods, it is often prescribed that there are three main methods, parsimony, likelihood based and distance based. Many practitioners swear by one method over the others (often for somewhat vague reasons), but in reality they often blend into each other and it is in understanding the interrelationships that a more robust understanding of phylogenetics emerges. One example of a bridging method is the Hadamard conjugation (Hendy and Penny 1993), which is a likelihood, distance and invariants method all at the same time (Waddell 1995, Swofford et al. 1996). In practice also, the methods have quite different and complimentary strengths. For example, parsimony is intuitive and has a useful exploratory flavor to it (especially when combined with software like Mesquite, http://mesquiteproject.org). Parsimony is also relatively fast for moderate sized data sets. Likelihood-based methods (such as maximum likelihood, ML, or marginal likelihood/Bayesian inference) offer detailed predictions of how the data/model should relate to each other. These are generally computationally expensive methods (Swofford et al. 1996), but tests of fit can be approximated (Waddell, Ota, and Penny 2009). Distance-based methods directly use the fundamental property of additivity of pairwise distances on the tree (Swofford et al. 1996). They also offer clear advantages in computational speed (or simply computability) on very large data sets. For example, a balanced minimum evolution (BME) criterion based method (e.g., Gascuel and Steel 2006) will now a complete an SPR cycle of tree search in $t^2$ time (where $t$ is the number of tips on the tree, Hordijk and Gascuel 2005). This is phenomenally fast since the number of sub-tree pruning and redrafting (SPR) tree rearrangements to be considered is itself $t^2$!

At present, in terms of fit of data to tree, there are few clear guides. Cladists have long held that measures such as the consistency index are a useful guide to how "clean" a data set is, but it is also well recognized that these are hard to calibrate for different sized data sets (e.g., Swofford et al. 1996). In terms of maximum likelihood on sequences, the likelihood ratio statistic may be used as a measure of fit, but for real phylogenetic data, the extreme sparseness of the data and complex parameter optimization and model (tree) selection makes it difficult to infer its expected distribution, except with simulations. More recently it has been shown it may have limited power to pick up deviations known to affect the tree selected as best (Waddell, Ota and Penny 2009). Distance methods, for example, generalized least squares, also report a type of likelihood ratio statistic (e.g., Bulmer 1991). None of these statistics report directly on the additivity of the data to the tree. Yet additivity is a key property of evolutionary models and is the basis of the consistency of nearly all distance-based methods. It would be appropriate to see it play a more prominent role in the evaluation of real analyses, rather than being hidden away in the fundamental mathematical proofs of criterion-based and algorithmic methods.

Another area that has received much attention over the past 20 years is how to infer the effect of errors on the tree. The sequence bootstrap (Felsenstein 1985, Penny and Hendy 1985) is one of the simplest, but it focuses mainly on stochastic error. Thus, for long sequences as used in phylogenomics, the bootstrap is pretty much guaranteed to report 100% support for all edges in the tree not sitting on a boundary (e.g., Ota et al. 2000). Concerns such as how to precisely calibrate these numbers in terms of more traditional test specific probability values (e.g., Shimodaira 2004), are therefore relegated to a secondary concern in phylogenomics. In theory, parametric Monte Carlo resampling (also called parametric bootstrapping) can better estimate stochastic error if the model is correct. However, the results are often not useful with real data where systematic error dominates. Bayesian methods often report posterior proportions, but these too focus on stochastic error. If this is the only source of error in the data, they are expected to be very accurate (when allowing for the selection of non-binary trees), but again, systematic error is ignored. Other methods such as Shimodaira and Hasegawa (1999) tests are sometimes invoked by more cautious practitioners, but they too principally measure stochastic error. They also involve a recentering step that causes them to be to too conservative when the model is close to correct, yet still fail to reject when the model is badly wrong (Waddell, Kishino and Ota 2000). By focusing on the stochastic resampling of the real data and not simply on the model predictions, it is possible to develop hybrid methods with some added robustness (e.g., Waddell, Kishino and Ota 2002), but ultimately, these too do not directly measure potentially serious systematic error.

For distance based methods there are three most promising approaches. Firstly, to use generalized least squares (GLS, e.g., Bulmer 1991, or better still iterated generalized least squares, e.g. Waddell 1995), which is close to ML methods for distance data. GLS is very computationally expensive even with time optimal algorithms, (being $O(t^4)$ for a tree, plus $O(t^6)$ for covariance matrix inversion, Bryant and Waddell 1998). Secondly there are flexi-weighted least squares methods (fWLS). Here the variance is assumed to be proportional to $d^P$, where $d$ is a distance and $P$ is a parameter that is either set by the user or learned (estimated) from the data (Felsenstein 1989, Sanjuán and Wróbel 2005, Waddell, Kishino and Ota 2007). The aim is to minimize the sum over all distances of $(d_{obs} - d_{exp})^2 / d_{exp}^P$, when $P$ is fixed. If the observed distance is used on the bottom line, then these methods are as fast as $O(t^3)$ for a single tree (Bryant and Waddell 1998). Results tend to suggest that variants of fWLS (e.g., Fitch and Margoliash 1967) can perform close to GLS in simulations (e.g., Gascuel 2000). Finally, there are balanced minimum evolution BME methods (e.g., Gascuel and Steel 2006). The algorithmic method Neighbor-Joining (NJ, Saitou and Nei 1987) has been explained as a heuristic approach to optimizing this criterion (Pauplin 2000). As mentioned above, algorithms such as fastBME (Desper and Gascuel 2002) now search faster, more explicitly and more accurately for the BME best fitting tree. It would clearly be very desirable to have more reliable methods for assessing the robustness of these types of trees.

Phylogenomics means different things to different people, but all too often it simply means making phylogenetic inferences with a large concatenated data set (e.g., D'Erhcia et al. 1996, Rokas et al. 2003). Some of the first prominent examples of phylogenomics were analyses of the mitochondrial genomes of mammals (e.g., D'Erchia et al. 1996). Here sequences were not so long that all edges automatically went to 100% bootstrap support, but it was clear that even in this context researchers were having trouble separating size of data set from quality of data set (Waddell, et al. 1999). Indeed, the first accurate classification of placental orders required the heavy discounting of many of the most prominent bootstrap results of mitogenomic analyses, such as hedgehogs at the root of placental mammals and rodents being noon-monophyletic (Waddell, Okada and Hasegawa 1999, Waddell, Kishino and Ota 2001). It was ironic that with such a history of issues surrounding the interpretation of bootstrap values, and their insensitivity to systematic error, that the concatenated bootstrap analysis of about 100 protein-coding genes from 8 yeast genomes was heralded as the answer to phylogenetic problems in the popular science publication Nature (Rokas, et al. 2003). The article attracted a variety of responses, often pointing out the fallacies of the articles key points (e.g., Holland et al. 2004, Phillips, Delsuc and Penny 2004). Nature clearly seems to have difficulty learning from history!

In this article we explore ways of measuring fit of distance data to a tree and then the use of the resampling of residuals to produce replicate data sets that more accurately reflect the total non-tree error in the data. The residuals represent the breakdown of additivity and comprise a mixture of total error, which is a mixture of stochastic or random sampling error, and systematic error. In the least squares context, (total error)$^2$ = (random error)$^2$ + (systematic error)$^2$ (e.g., Waddell et al. 1994, Waddell 1995). The former goes to zero with long sequences, while the later remains near constant. It is shown that residual resampling is a viable alternative to the sequence-based bootstrap for phylogenetic analysis. Residual resampling is a robust method that should be considered by data analysts and reported to readers as one of a number of different assessments of the robustness of the data. Application to both fWLS and BME is illustrated and software for the use of RR is made available.

## 2 Materials and Methods

The yeast alignment of 107 protein-coding genes from 8 species with a total length of 127,026 nucleotide sites is from Rokas et al. (2003). These were translated to amino acid sequences using the universal nuclear code.

The constrained ML tree for nucleotides under the general time reversible (GTR)

invariant sites plus Γ model was found using successive iterations of tree search then repotimizing the parameters on the best tree until no further improvement was found using PAUP4.0 (Swofford 2000). All nucleotide distances were calculated using PAUP. A test version of PAUP 4.0 (beta 105 or 106) was also used to find optimal ML DNA and amino acid trees along with their parameters, and to perform all other tree calculations including least squares with any $P$ value. Amino acid distances we calculated using PAUP (p-distance, Swofford 2000), Monty (GTR distance, Waddell and Day unpublished) or PAML (ML distance, Yang 1999). Note, all parametric analyses assumed a GTR + Γ model. While ML + Γ + $p_{inv}$ models (Waddell and Penny 1996, Waddell and Steel 1997) often tend to fit real data better, their use is precluded since not all the programs offer genetic distances with such site rate distributions.

Parametric "replicate" sequence data sets were generated using Seq-gen (Rambaut and Grassly 1997). PERL scripts were used to automate and integrate tasks, including the resampling of residuals that were then feed to other programs to calculate distances and then to PAUP for tree search. *Q-Q* plots and Shapiro Wilk tests were made with Matlab 7.40.

## 3 Results

As a preliminary step, the GTR + Γ model was fitted to the yeast DNA sequence data. It yielded the ML weighted tree:
((((((Scer:0.06121848,Spar:0.03686066):0.03380840,Smik:0.08252218):0.03918712,Skud:0.08201029):0.03456079,Sbay:0.07943537):0.25106293,Scas:0.37869060):0.15146443,Sklu:0.31554060,Calb:1.49403231),
which is a tree of the same topology as that in figure 9, row 2, with base frequencies A:0.319405 C:0.176220 G:0.198105 T:0.306270, shape parameter 0.415853, and relative rate matrix entries AC: 3.0576803, AG: 8.5141266, AT: 1.2600993, CG: 2.676979, CT: 14.855965 and GT: 1.

**3.1 Flexi-Weighted Least Squares Criteria**

Least squares fitting was initially implemented in tree selection by Cavalli-Sforza and Edwards (1967) using ordinary or unweighted least squares and Fitch and Margoliash (1967) with weighting proportional to the observed distance squared ($d_{obs}^2$). It was later implemented in a more general form such that the variance was assumed to be proportional to $d_{obs}^P$, where $P$ was a parameter specified by the user (e.g., the FITCH program within PHYLIP, Felsenstein 1989). In order to select a suitable value of $P$, users might seek to minimize the sum of squares, but this would not work because the sum of squares was not on a fixed scale (since the weights change with $P$). One solution was to aim to minimize a scaled measure of fit such as the percent standard deviation (%SD) of the observed distances mapped back to the tree (e.g. Waddell and Kishino 2000). Unfortunately, the equation being used in PAUP at that time was not calculating this correctly. A formula to work with any power of $P$ was presented in Waddell, Kishino and Ota (2007), which is

$$\%\text{SD} = \left(\frac{N}{\sum_{i=1}^{N} d_{obs_i}}\right)^{1-\frac{P}{2}} \left(\frac{1}{N}\sum_{i=1}^{N} \frac{\left(d_{obs_i} - d_{\exp_i}\right)^2}{d^P_{obs_i}}\right)^{\frac{1}{2}} \times 100\% \quad (1)$$

(Note, there is a typo in Waddell, Kishino and Ota 2007, where the first "N" in the equation above is printed as "1").

An alternative approach is to select *P* based on the principal of maximum likelihood, for example maximizing equation A8 of Waddell, Kishino and Ota (2007) (see also Sanjuán and Wróbel 2005 for an independent derivation of likelihoods on weighted least squares trees). It is striking that the formulae for the likelihood and the percent standard deviation are the same except for the form of the rescaling mean being used (arithmetic with percent deviation or geometric with likelihood, Waddell, Kishino and Ota 2007). Note that the likelihood derived is itself a type of likelihood ratio statistic; it is proportional to the change in likelihood from the highest possible likelihood for that data, which here would have sum of squares zero. Ignoring constants and typos, Equation A8 of Waddell, Kishino and Ota (2007) can be rewritten

$$1/\text{Likelihood(per distance } d) \sim \left(\left[\prod_{i=1}^{N} d_{obs_i}\right]^{1/N}\right)^{P/2} \left(\frac{1}{N}\sum_{i=1}^{N} \frac{\left(d_{obs_i} - d_{\exp_i}\right)^2}{d^P_{obs_i}}\right)^{\frac{1}{2}} \quad (2)$$

Thus, the aim is to minimize equation 2 to maximize the likelihood.

The percent standard deviation approach in equation 1 differs from the likelihood by also rescaling by the inverse of the mean distance. This is desirable when comparing distance data sets without an external estimate of the variance, since multiplying all the distances by an arbitrary factor such as two will change the likelihood, but not the percent standard deviation. This scale invariant property can be imposed by multiplying the likelihood by the inverse of the geometric mean of the distances, to give

$$\text{Geometric }\%\text{SD} = \left(\left[\prod_{i=1}^{N} d_{obs_i}\right]^{1/N}\right)^{P/2-1} \left(\frac{1}{N}\sum_{i=1}^{N} \frac{\left(d_{obs_i} - d_{\exp_i}\right)^2}{d^P_{obs_i}}\right)^{\frac{1}{2}} \times 100\% \quad (3)$$

Equation 3 is proportional to the likelihood of the data, invariant to arbitrary changes in the scale of the distances, and intuitively very close to the %SD measure of additivity. We abbreviate the geometric mean based %SD like measure of equation 3 to g%SD.

The inverse of the arithmetic mean might also be combined with equation 2 to give an ag%SD measure, and it too is considered further below. Note that these equations will identify the same optimal parameters as likelihood within a distance data set, but will differ when the distance data set itself is varying. The composite criterion (equation 3) is valuing absolute precision over arbitrarily changing the likelihood by, for example, multiplying all distances by a factor of two. Equations 1 and 3 are acting as direct measures of the additivity of distances to a tree (minimizing root mean square deviations). If we know the edge lengths of the true tree, they are also a metric for measuring how far we were from it. In practice, distance tree building methods assume that the (weighted) tree the distances are most close to additive upon is the best estimate of the true tree.

Note that changes in equations 1 and 3 can be converted back into log likelihood ratios. Using g%SD as an example, firstly, undo the normalization by multiplying by the geometric mean, take the log of this value, then multiply by the number of distances in the data ($t(t-1)/2$), where $t$ is the number of tips in the tree. Thus, a small difference in the g%SD per distance can translate to a significant difference (e.g., two or more) in the log likelihood difference of the whole data set, especially on larger trees with many tips.

As a general guide, it is useful to note that most DNA and protein distances have at their core a transformed multinomial sample. At low rates of substitution, the transformation is close to unity. Also, at low rates of substitution and long sequences, a multinomial distance behaves like a simpler binomial p-distance. The variance of a p-distance at low rates of change is effectively proportional to its size. Under such conditions, and flexi weighed least squares, the optimal value of parameter $P$ is close to one, and increasing stochastic and systematic error (which affect larger distances proportionally more) pushes this upward towards 2 or more.

Another useful point to remember is that distance estimation and additivity have an interesting relationship to each other. When the generating tree is clock-like, and the process of evolution stationary, then pretty much any evolutionary distance estimator will be monotone increasing and hence also additive on the same unweighted tree (e.g., Waddell and Steel 1996 contains a proof of the monotone increasing nature of distances for any distribution of site rates for any stationary model). Here, monotonicity preserves the consistency (and observed additivity) of pretty much any model based-distance and distance-based tree inference method (e.g., Jukes-Cantor distances and UPGMA) under a clock and stationarity, despite the fact that these same conditions can lead to the inconsistency of ML (Waddell 1995)! However, mismatch between generating model and reconstruction model also guarantees systematic error on estimates of edge lengths (which equals internal node heights, under a clock). This may be called the "accordion effect" since the tree/distance combination itself can expand (the root gets proportionately deeper) or contract (the root gets proportionately closer), rather like an accordion, with minimal effect on observed additivity. Accordingly, equation 3, for example, will, in expectation, not detect any mismatch of estimated and tree distances in such situations. By extension, equation 3 may have reduced power to detect deviations when the generating tree is near a clock and violations of stationarity are not too strong. Even away from the clock, if errors are additive on the edges of the tree, then such errors are readily hidden by the adjustment of edge length values (Gascuel 2000).

**3.2 Fit of Nucleotide Sequences**

We begin our exploration of the fit of data using flexi-Weighted Least Squares (fWLS) by fitting of nucleotide sequence distances derived from the yeast data set. Figure 1 shows the fit according to equation 1 and 3 for various values of *P* and three different genetic distances. The distances being considered are the Hamming, edit, or p-distance (the last name being widely used in molecular systematics, and having no relationship to parameter *P*), a GTR + Γ ML distance and a GTR + Γ pairwise distance. The parameters of the ML GTR distance (referred to as the ML distance below) are set to those found for the same model applied to the sequence data. The free parameters were not explicitly optimized according to the fWLS likelihood (equation 2), nor do they aim to maximize either equation 1 or 3 as measures of additivity. The GTR pairwise distance (referred to the GTR distance below) is a ML distance based on the marginal F matrix of nucleotide changes between each pair of species, with the Γ shape parameter alone coming from the ML sequence-based estimates (Waddell and Steel 1997). For each value of *P* a full TBR tree search was performed until completion, with each tree being reoptimized with the constraint that all edge lengths are positive (via an extension of the fast least squares algorithms described in Bryant and Waddell 1998 implemented in PAUP4.0).

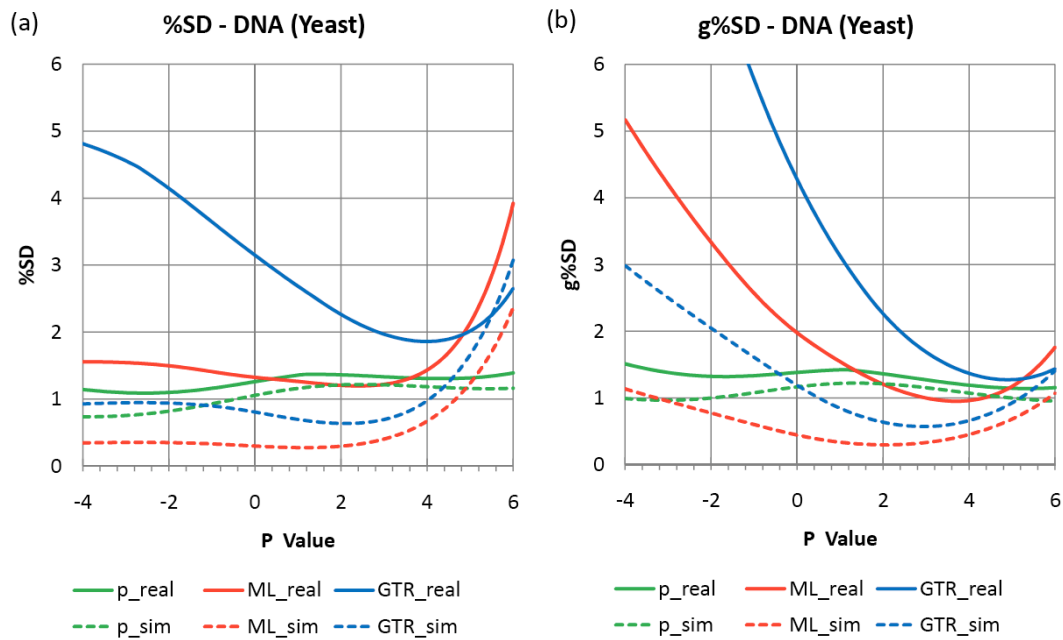

Figure 1 (a). A plot of the arithmetically adjusted percent standard deviation, %SD (equation 1) fit against parameter *P* for the yeast nucleotide data (*P* was in steps of 0.1). Three distances are used, the Hamming or p-distance, an ML distance, and the pairwise GTR distance. These distance transforms are applied to both the real data and a set of simulated sequence data generated under the parameters of the ML GTR + Γ model fitted to the real sequence data. At each point a full TBR tree search with simultaneous reoptimization of all edge lengths was performed. Figure (b). Identical to (a) except that the geometrically adjusted percent standard deviation, g%SD (equation 3) measure of fit is used.

Figure 1a shows some interesting features. Prominent is the fact that the GTR distance does not fit the tree especially well in terms of additivity. Indeed, it appears to fit considerably worse than the p or ML distances. The cause is not immediately clear. This is partly a stochastic effect; that is, pairwise GTR distances have larger sampling errors than sequence based ML distances asymptotically under the model (Waddell and Steel 1997). The plot of simulated data in figure 1a clearly shows there is a difference in the purely stochastic effects between GTR and ML distances. However, with the real data it appears there might be something more going on as the gap is even wider than that between ML and GTR distances under the model. It is possible that the worse fit of the pairwise GTR distances compared to the sequence based ML distance on the real data is due to more pronounced effects of long edge attraction (Felsenstein 1978), and just as, or perhaps, more importantly, long edge repulsion (Waddell 1995), with these effects being powered particularly by non-stationary base frequencies. Long edge repulsion occurs when base composition differences makes the F matrix appear less similar than would be the case with stationary frequencies and since there are logarithmic transformations and asymptotes involved, this can have a larger numerical effect than long edge attraction.

In figure 1a the ML distance fits slightly worse than the p-distance and at different optimal values of *P*. The ML distance is clearly fitting much worse than expected under the model they were generated under. The p-distance in contrast fits not much worse than under the model; but the p-distance is never expected to be perfectly additive. Of all the distances, p-distance mathematically must show the least stochastically driven departure from additivity, but is always expected to show systematic deviation from additivity, except at zero rates of change. It also shows multiple optima in equation 3, with the deepest minimum visible at *P*~-3. The low value at which *P* is optimal is quite unlike the expected relationship between an additive nucleotide distance and its model-based variance, so this serves as a warning sign. If *P* is limited to the range of values that could realistically mimic the relationship of an additive sequence distance to its model based variance (i.e. *P* > 0) then the ML distance appears to be slightly more additive than the p-distance.

Figure 1b shows the fit as measured by the likelihood criterion reweighted by the inverse of the geometric mean, that is g%SD (equation 3). Whereas we see the fit functions in figure 1a spilling out to the left, in figure 1b the curve seems more symmetric when there is a single minimum. The ML distance is now fitting best of the three distances considered. However, the ML distances from the real data are clearly a much worse fit, by a factor of about four, from that expected under the model as shown by the curve for the simulated distance data. This is clearly not a good sign. The sequence length of this data (over 100,000 bases) is long enough to drive the expected non-additivity down to about ¼ of a percent under the model, as shown by the simulated data. Another way of looking at this is that the actual least squares error is about $4^2$ or 16 times as large as it should be. Thus, if stochastic error were the only cause, the effective sequence length would be only 1/16 or so of what it is assumed to be! (The sum of squares due to stochastic error will, asymptotically, decrease linearly with the length of the sequence). Thus, the use of g%SD allows us to estimate the effective length of the yeast sequence data, which appears to be less than 10,000 bases!

Figure 2a shows the use of the arithmetic instead of the geometric mean in adjusting equation 2. Note that by using the geometric mean, the overall fit is exactly the same at *P* = 2 as estimated by equation 3. This in turn tends to see g%SD closer in value to %SD than ag%SD in the range of $P = 1$ to 3, which is where the optimal value is expected to typically lie for distances derived under simple Markov stochastic sequence models. Figure 2b shows the plot of equation 2, which is inversely proportional to the likelihood (so minimizing it increases the likelihood within a dataset). All the plots in figures 1 and 2 can be interconverted by knowing the arithmetic and geometric means of the distance data sets. For the yeast data these are: p-distance real DNA (arithmetic mean 0.25810, geometric mean 0.23514), ML-distance real DNA (0.95515, 0.63997), and GTR real DNA (0.83658, 0.61547). For the simulated DNA data they were p-distance (arithmetic mean 0.27016, geometric mean 0.24464), ML-distance (1.00263, 0.67426), and GTR (0.99534, 0.67393).

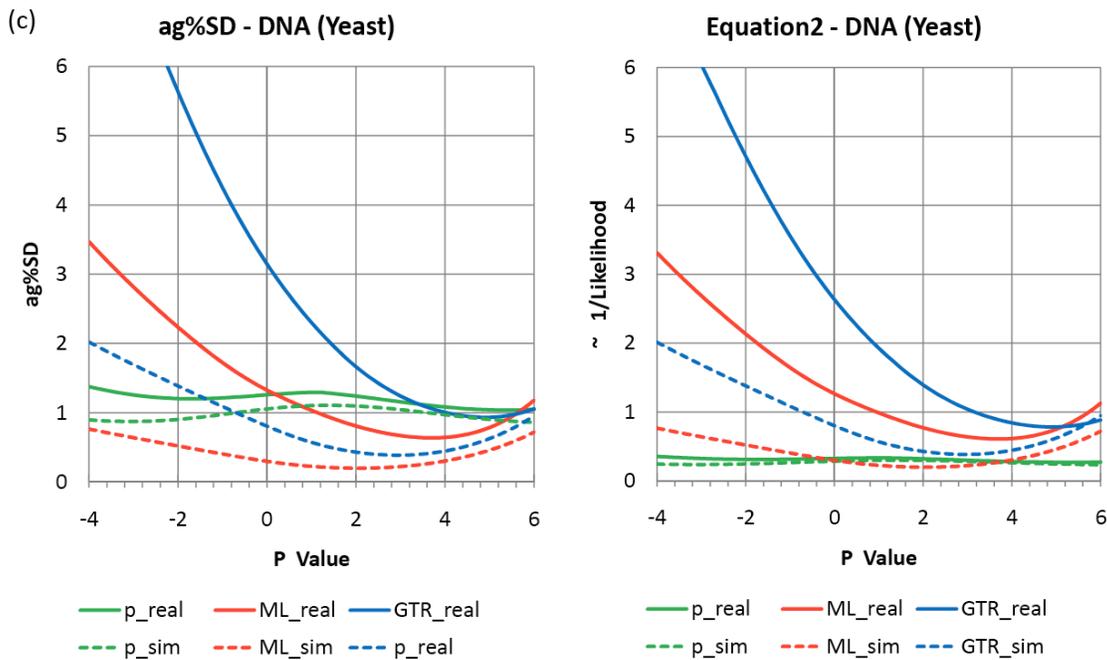

Figure 2 (a). A plot of equation 2 multiplied by the inverse of the arithmetic mean, which we call ag%SD against parameter $P$ for the yeast nucleotide data. The distances and calculations are as in figure 1. Figure 2(a) is a translation of figure 1(b) that matches figure 1(a) less exactly near the likelihood minima. Figure (b). Identical to (a) except that equation 2, which is proportional to the inverse of the likelihood, is plotted. The problem with this measure is that if the distances are arbitrarily multiplied by scalar $x$, equation 2 changes by a factor of $x$ and the likelihood changes by a factor of $1/x$. Thus, it is difficult to compare datasets to each other.

**3.3 Fit of Amino Acid Sequences**

We now consider how well the amino acid based distances estimated from the yeast data fit to a tree. Figure 3a and b clearly shows that the additivity of amino acid distances measured by %SD or g%SD is approximately twice as good as that encountered above with nucleotide distances. Figure 3a clearly shows the deficiency of using the arithmetic and not the geometric mean, so the rest of this discussion will focus on g%SD. While the shapes of the curves look different, again, since the minima occur relatively near $P = 2$, the fit values given by equations 1 and 3 at the minima of equation 3, are close. The p-distance is once again the best fitting, but this time it shows a single minimum. Indeed, the curves for all distances look much more regular and the optimal $P$ value is close to 2 in all cases. An optimal $P$ of near 2 suggests that the total error rises linearly with the evolutionary distance.

The simulated amino acid data fits slightly worse than the nucleotide data (figure 1a), as the stochastic error of amino acid sequences will be higher (there are three nucleotides for every amino acid). This combined with the higher additivity observed with the real data, sees the difference in additivity between real data and model expectations reduced to a factor of two or less from when using nucleotides. Thus, the amino acid distances generally fit model expectations better than with nucleotides. Notice also that the exact value of $P$ at which the minimum occurs varies between ~1 and 2.5 one this small sample of simulated data.

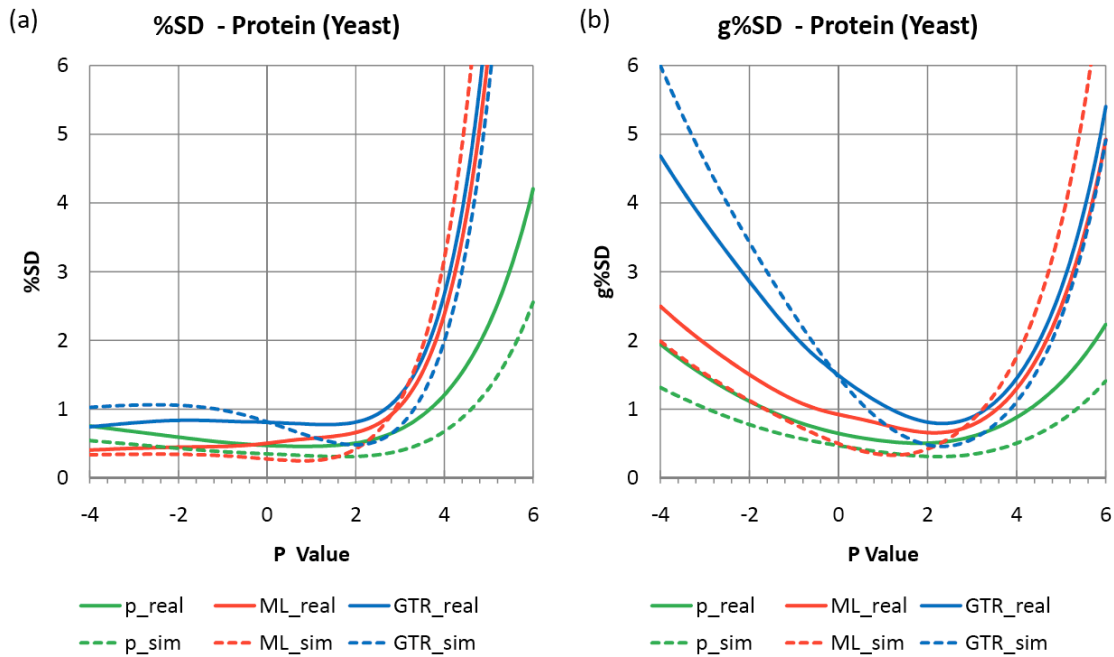

Figure 3 (a). A plot of the arithmetically adjusted percent standard deviation, %SD (equation 1) fit against parameter $P$ for the yeast protein data. Three distances are used, the Hamming or p-distance, an ML distance, and the pairwise GTR distance. These are applied to both the real distances and a set of distances generated under the parameters of the ML GTR + Γ model fitted to the amino acid sequence data. At each point a full TBR tree search with simultaneous reoptimization of all edge lengths was performed. Figure (b). Identical to (a) except that the geometrically adjusted percent standard deviation, g%SD (equation 3) measure of fit is used.

Figure 4 shows the relative fit of the various distance measures to a fWLS tree. The real nucleotide distances show the worst fit, with the one interloper being the simulated p-distances. These are distances are showing clear signs of either suffering from systematic error or else having much larger stochastic error than expected. This latter hypothesis can be tested by randomly selecting half the Rokas data (either half the genes or every second codon). If the only source of error is stochastic, then both of these treatments should increase the g%SD$^2$ by about 2. The real amino acid distances fit markedly better than the real nucleotide distances and the simulated data tends to fit best of all, with the GTR distances showing worse fit than the ML distances, probably due to being more susceptible to stochastic error.

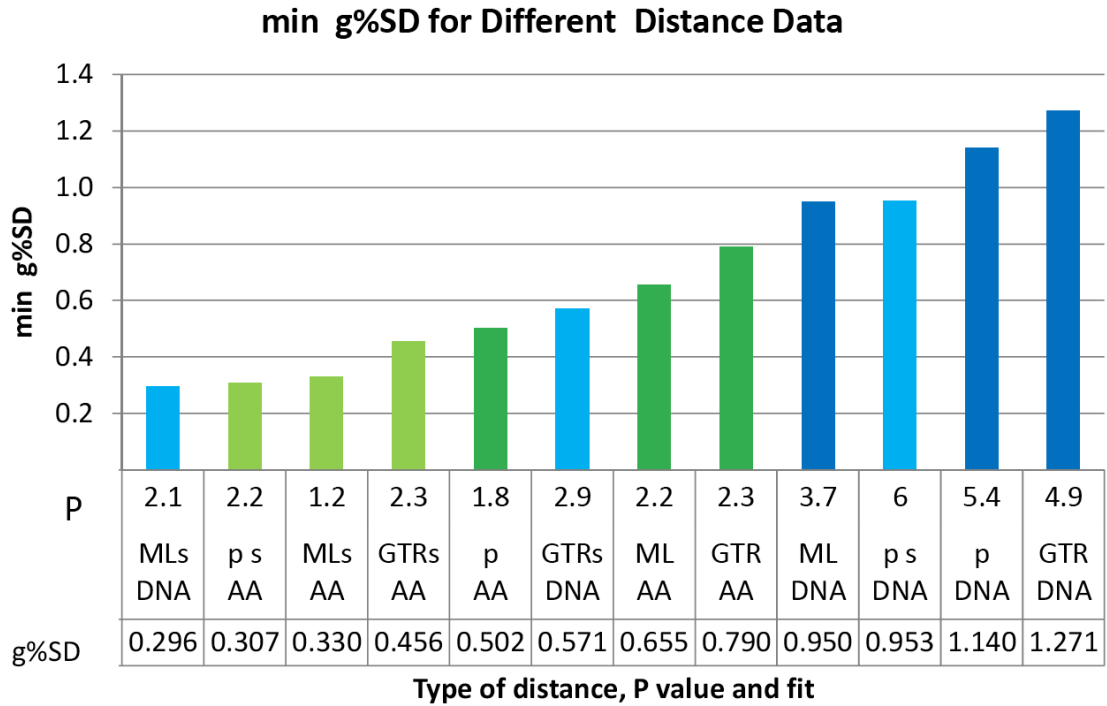

Figure 4. A plot of the optimal fit by equation 3, g%SD, for six distance transformations of the DNA (blue) and protein (green) data, for both real and simulated data (indicated by lighter colors and an "s" in the distance name). Also indicated is the value of *P* at which the optimal fit occurred.

While it is difficult to directly compare the likelihood of nucleotide to amino acid sequence data, it seems clear that in this instance the amino acid data are more additive on a tree. Since additivity is an essential property of substitution-based genetic distances, it seems useful to consider its fit in deciding which distance transformation to use when estimating a genetic tree. Since least squares additivity measures both stochastic and systematic error, it seems a natural arbiter in deciding how complex a model to use when searching for a tree. This may resolve the conundrum seen in simulations that the use of simpler models, or even the p-distance, often does a better job for estimating the tree than the true model (e.g., Kuhner and Felsenstein 1994). Obviously, this cannot be true with large amounts of data (since only a sufficiently parameterized model can be full additive), so the question is how to judge the balance. Here we suggest minimizing equation 3 may act to fulfill this role.

For the amino acid distances based on the real protein sequences the means were p-distance (arithmetic mean 0.21216, geometric mean 0.15464), ML-distance (0.54063, 0.29422), and GTR (0.48922, 0.26444). For the simulated protein data they were p-distance (arithmetic

mean 0.21805, geometric mean 0.16213), ML-distance (0.50197, 0.27721), and GTR (0.51206, 0.28410).

**3.4 Residuals**

It is as important to look at the residuals when fitting distances to a tree as when fitting points to a line. In these examples the raw residuals are $d_{obs} - d_{exp}$, and the standardized residuals are proportional to $(d_{obs} - d_{exp})/(d_{obs}^P)^{0.5}$ (which is sufficient for measuring the linearity of a Q-Q plot, as done later). Figure 5 shows a plot of the standardized residuals for the ML amino acid distances at the optimal $P$ value and on the best fitting tree. The raw residuals are clearly getting larger with the distance. The standardized residuals are more randomly distributed, but the real residuals are clearly larger than those of the simulated data.

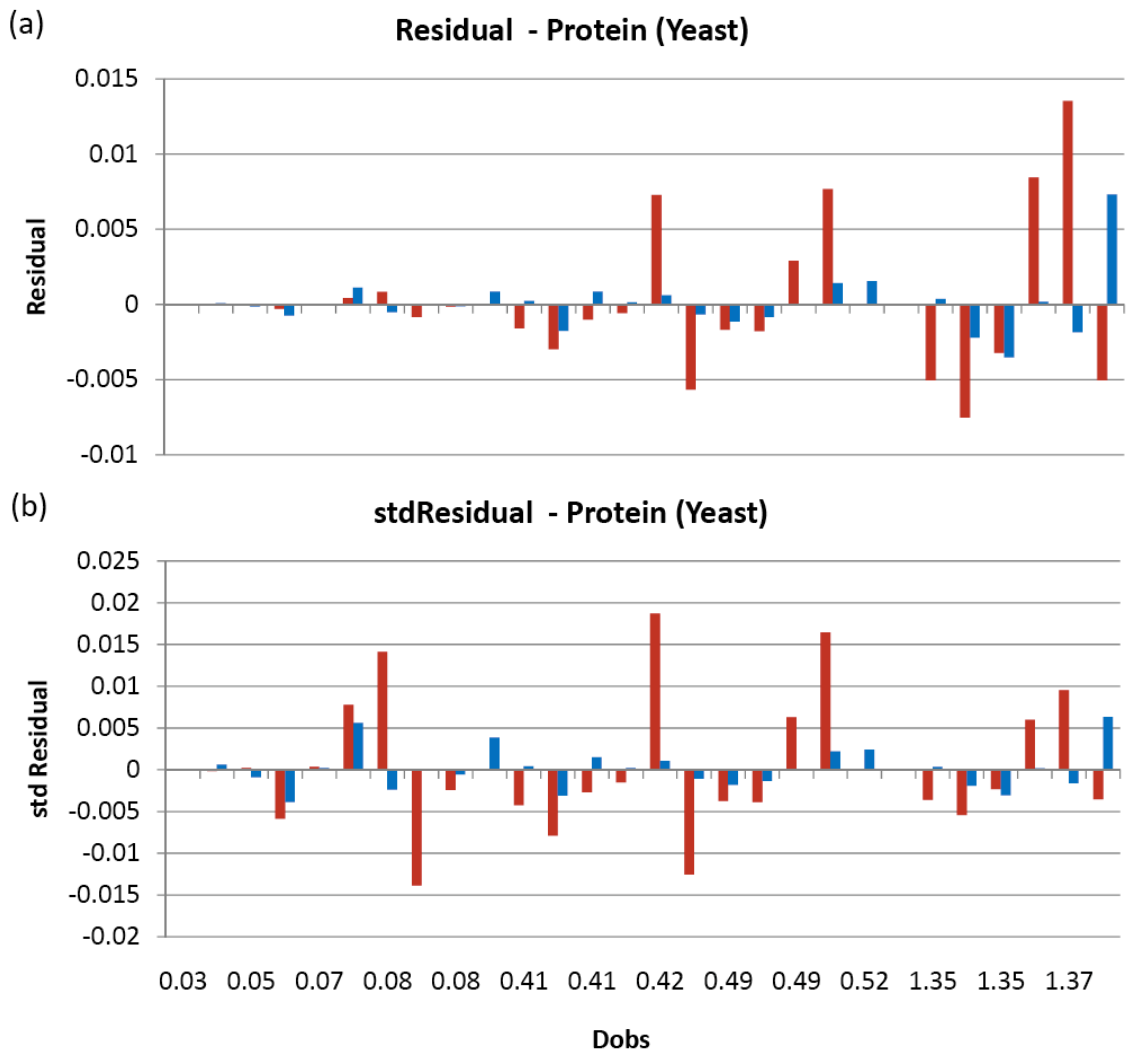

Figure 5. A plot of the raw (a), $(d_{obs} - d_{exp})$, and standardized (b), $(d_{obs} - d_{exp})/(d_{obs}^P)^{0.5}$, residuals for the amino acid ML distances, plotted against the observed distance, $d_{obs}$. The red are the residuals from the real data ($P = 2.2$), while the blue are from the simulated data ($P=1.2$). While the standardized residuals of the real data look far larger, and their sum of squares differ by a factor of 0.001616 to 0.000155 or about 10.4,

while their root mean squares differ by a ratio of about 3.22, the g%SD difference is almost exactly a factor of two (0.655 vs. 0.330). This is because the geometric correction term in this equation is $(0.29422)^{2.2/2-1} = 0.885$ for the real data while for the simulated data it is $(0.27721)^{1.2/2-1} \sim 1.671$.

Another useful way of looking at residuals is with quantile-quantile or Q-Q plots. Figure 6 shows a Q-Q plot of the raw and the standardized residuals for amino-acid distances. The raw residuals are clearly not fitting with the expected quantiles of a normal distribution, but after standardization they look reasonable. There are a number of tests for the normality of i.i.d. residuals and one of the most popular and powerful of these is the test of Shapiro and Wilk (1965). In this instance it clearly rejects the normality of the unstandardized residuals, but is borderline for rejecting the standardized residuals. This illustrates the point that proper weighting of residuals is important when fitting evolutionary trees to distances from real sequence data.

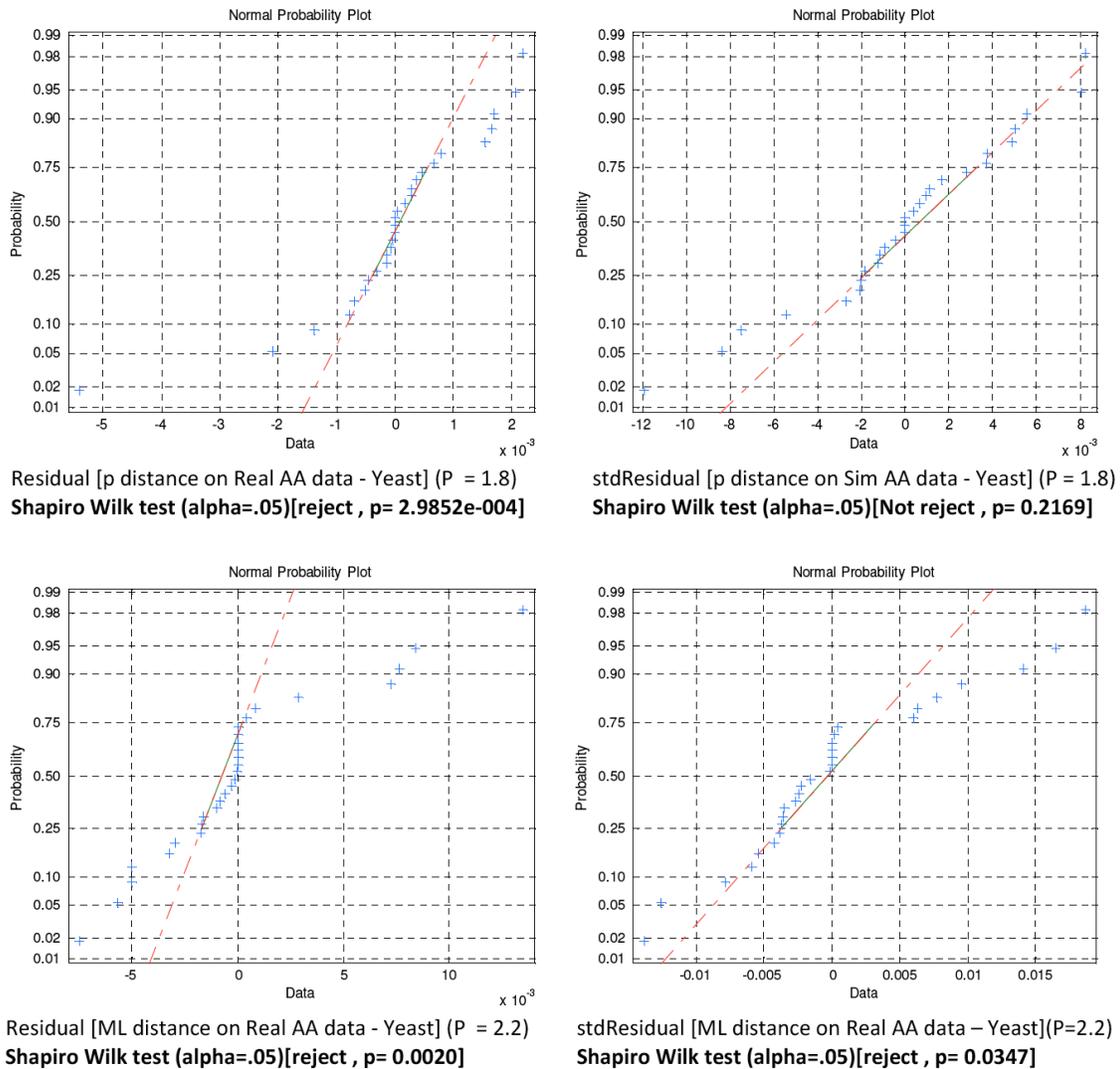

Figure 6. Q-Q plots and Shapiro Wilk test results for the raw (a), $(d_{obs} - d_{exp})$, and standardized (b), $(d_{obs} - d_{exp})/(d_{obs}^P)^{0.5}$, residuals for the amino acid p and ML distances.

Before we leave this topic, it is useful to check how well the residuals of distances from sequence data follow an i.i.d. normal distribution. There are some results that suggest weighed least squares results can closely approximate generalized least squares results (Sanjuán and B. Wróbel 2005). This is despite potentially strong correlations of errors on distances due to the evolution of sequences following a tree, and thus there may be many distances sharing the bulk of their information from the same sequences evolving along the same (internal) path in the tree. As Figure 7 shows, the Shapiro Wilk test is not robust in this situation and rejects too often. Given the sequence sample size, it is almost certain the distances are correlated multivariate normal (e.g., Waddell et al. 1994). The marginals will be very close to normal, but the correlations are resulting in clustering of results, such that the i.i.d. based Shapiro Wilk test statistic is rejecting to often in this instance (e.g., on the Rokas data an Shapiro Wilk test result of 0.05 has real probability of ~0.2). The results are even worse for larger trees. It is possible to attempt to correct for correlation structure by multiplying the observed residuals by the inverse of the root of the variance-covariance matrix of the distances. A simple form of the variance covariance matrix of the distances may be estimated from the optimal weighted tree, by assuming the observed distances were generated on it and have additive errors. Later we look at residuals of BME and NJ, and in this case figure 7 shows their residuals fit i.i.d. expectations even worse.

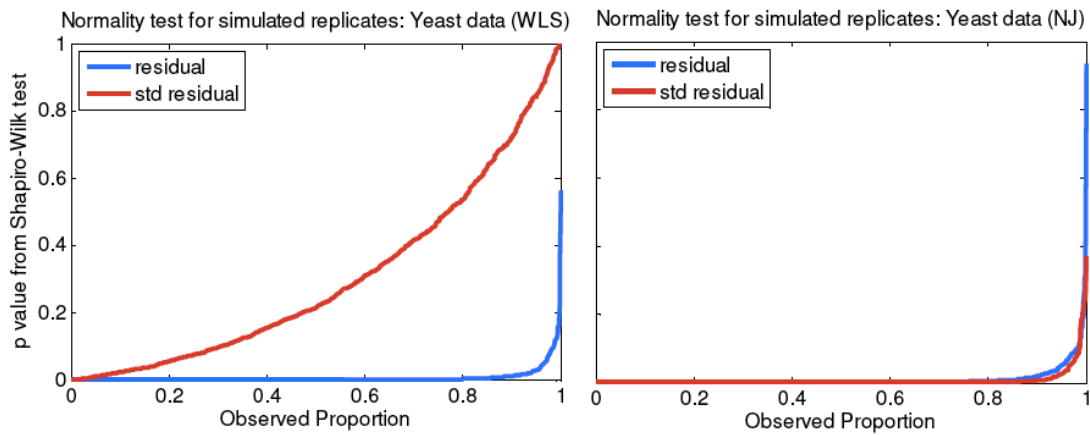

Figure 7. Simulated results of the Shapiro Wilk test made upon ML distance residuals from DNA sequences simulated according to the ML GTR + Γ parameters of the original yeast data. On the left are residuals from fWLS and on the right from NJ.

Finally, note the observed residuals are shrunk or deflated from their expected size by the act of fitting a tree and also parameter $P$. Without knowing the generating model, it is impossible to estimate how much they have shrunk by, but a broad estimate can be made. If it is assumed that the data fit the fWLS model and are away from boundaries in the parameter space (e.g. Ota et al. 2000), and assuming i.i.d. residuals, then the total sum of squares should follow a chi-square distribution with degrees of freedom $x$, where $x$ are the number of distances (here $t(t-1)/2$, where $t$ is the number of tips on the tree). The expected value of a chi-square distribution is its degrees of freedom. Under the model, each fitted parameter is expected to reduce the degrees of freedom by 1, and there are $2t-3$ fitted edge lengths, plus parameter $P$. Thus, with a binary tree, the fWLS sum of squares is estimated to have shrunk by a factor of $(t(t-1)/2-2t-3-1)/(t(t-1)/2)$ or $(t^2 - 5t - 8)/(t^2 - t)$. For the Rokas data set, this implies that the sum of squares is only about 50% of what it is from the true model (ignoring the effect of tree search, a form of model selection). As the data set gets larger, this shrinkage effect is less pronounced; for example, with 40 taxa the shrinkage is estimated at ~10% and by 100 taxa about 4%. There is also shrinkage occurring because of tree search. The impact of this factor can be made with simulations.

**3.5 Resampling Residuals with Replacement**

In this section we resample residuals and add them back to the distance data as a way of measuring the effect of non-additivity, whatever its cause, on the tree. This non-additivity error is a measure of both stochastic and systematic error. Recall, the second component does not necessarily go to zero as sequence lengths go to infinity. For these reasons, resampling of residuals will often be more relevant to phylogenetic practice than the sequence bootstrap. In addition, the standard sequence bootstrap is based on an i.i.d. assumption; if there are correlations amongst sites, then this too will cause the sequence bootstrap to err towards being too optmistic, but it will not affect the resampling of residuals.

Resampling and bootstrap-like approaches can be done in different ways and it is useful to set these out as decisions. One issue is whether to add the resampled residuals to the original distances, which already contain a degree of non-additivity, or to add them to the expected distances of, for example, the optimal tree? If you are confident the model is really a tree and that you have a reasonable estimate of the correct tree, you can logically argue that the expected distances should be used. Alternatively, if you do not want to assume that the distances come from a tree or a single tree, or that you do not have a good proxy of that tree, you may instead argue that adding error onto the observed distances is a more appropriate approach. Adding error to the observed distances carries the danger of creating over dispersed resampled estimates. However, if the "fit" gap between simulated and real data is significant and substantial, then using the expected distances carries its own risks.

Typically, we do not know the error between the "true" or generating tree and the observed data, but we do know that it is almost always underestimated due to extensive tree search (model selection) as well as the fitting of free parameters on each model. Compensation for this can be done in a number of ways, one of the most simple being the inverse of the deflation factor described in the last section (which can be called inflation or reinflation of the residuals). However, there is often a significant probability of underestimating the size of the true non-additivity (measured for example, by g%SD of the observed distances to the true tree and its edge lengths). In situations where the true scale of uncertainty is itself poorly understood, then a more conservative approach is often warranted in practice. In reality, it is not unreasonable to present the reader both estimates when there is reasonable doubt as to the nature of the true model and the real distance from it.

Note that it is the standardized residuals that are sampled randomly with replacement and the actual residual to add back to the observed or the expected distance (depending n which is being used) is that residual multiplied by $d_{exp}^P$. This should be done at the optimal value of $P$ for the simulated data. Since this may be very expensive to calculate, as an approximation, it may be fixed to the optimal $P$ value for the original data. Rather like imposing logical boundaries in the tree search space (such as disallowing negative edge lengths) the user may also chose to constrain $P$ to the range expected for a Markov process on a tree, which is $P$ approximately 0 to 4, with 1 to 3 seeming more likely.

These four resampling possibilities discussed are applied and results are shown in figure 8 for each of the distance measures on both simulated and real yeast distances. In this example, the addition of reinflated residuals to the expected distances does the best job of mimicking the degree of observed non-additivity seen in the original analysis (the true non-additivity is much larger for the p-distance, for example, but the accordion effect is hiding much of this).

When resampling residuals in a standard linear regression the question of leverage arises. That is, some distances have their residuals shrunk more than others since they contribute a disproportionate amount to the total sum of squares. One way this might occur is if trees are estimated from repeated distance matrices (e.g. multiple genes) in which case the distance replicate most different to the mean distance will have most leverage. It may also occur since some paths in the tree are estimated by more distances than others. This topic requires further study as applied to resampling the residuals of trees.

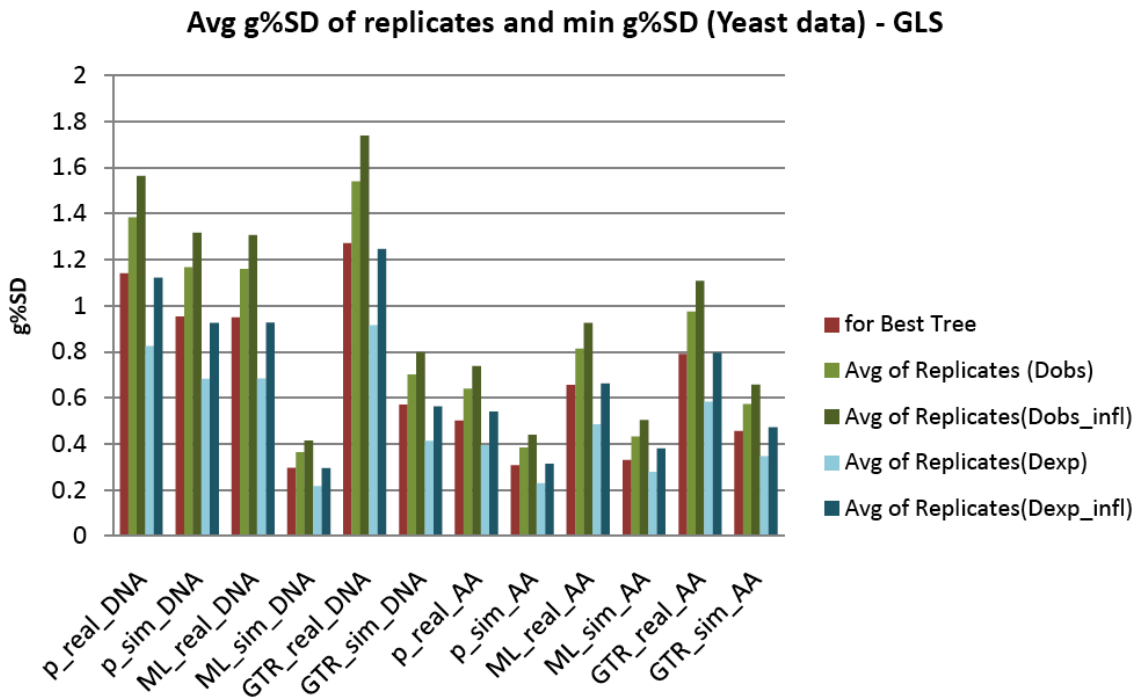

Figure 8. Value of g%SD (equation 3) with resampling of residuals on yeast distance data sets under different sets of assumptions based on 1000 replicates. The red is the original fit, the light green the mean of the replicates adding residuals to $d_{obs}$, the dark green adds reinflating the residuals, the light blue is the mean of the replicates with residuals added to $d_{exp}$, and dark blue, also reinflating the residuals.

### 3.6 Resampling Residuals (RR) and Support Consensus Trees

One of the issues with analyzing very long sequences is that the character resampling bootstrap will give 100% support to all edges of the tree, except in unusual situations where parameters are right on the boundary of the parameter space (e.g. an internal edge is exactly zero length). Here we show that the resampling of residuals approach does not necessarily lead to 100% bootstrap support even when the usual bootstrap does.

Figure 9, row 1, shows the consensus tree of the p-distance nucleotide-based distance tree after resampling of residuals with replacement. Note that there are two short edges that remain uncertain in that they do not occur 100% of the time in all trees. The second row shows the support on edges of the ML DNA based distances. The tree has changed, but there remains some uncertainty in these two parts of the tree. Figure 9, row 3, shows support on edges of the ML amino acid based tree. Note that the inferred tree remains that identified by sequence-based ML, and the support is now close to 100% on all edges. Note, the even though all edges are now have close to 100% support on all edges, this does not necessarily mean that residual resampling is behaving the same a the sequence bootstrap. If we measure the actual fluctuations in the edge lengths, as is necessary when estimating the error on divergence time estimates, then RR is still showing over ten times more variance in edge lengths than the i.i.d. sequence bootstrap.

As mentioned above, the least squares fit of tree for real relative to simulated data allows us to estimate the effective sequence length of the data assuming that the only reason for misfit is stochastic error (obviously much larger than under i.i.d. assumptions due to positive correlations between sites). Another way of looking at the same thing, is to ask how few of the 100,000 plus sites in the original yeast data need to be bootstrapped (aka Felsenstein 1985 or Penny and Hendy 1985) in order to show the same level of uncertainly as the residual resampling trees. With a p-distance and DNA data, but limiting the bootstrap resampling to 8000 characters, the support for the clade *S. kud* and *S. bay* drops to a little lower than RR (figure 9, row 1) at 71% vs. 74%, but the support for the clade excluding just *C. alb* and *S. klu* is much higher 99% versus 81%. Support for this second clade does not become comparable under i.i.d. assumptions until just 1600 characters are resampled (at which point the support for the clade *S. kud* and *S. bay* has dropped to 59%). Thus, resampling residuals suggests that the effective sequence length of this data when using fWLS with the p-distance is in the range of 1500 to 8000 independent nucleotides or just 1.2-6.3% of the apparent sequence length!

Thus a more detailed analysis of this genomic data suggests that far from having the apparent resolving power of over 100,000 characters, genomic data may really be behaving like a few thousand "good" characters, which is in good agreement with the estimate made earlier from the size of the sum of squares compared to simulated data. Further, even if the apparent sequence length were to go to millions of sites, it is unlikely to change the effective sequence length by much since the expectation here is that the majority of the difference between RR and the sequence bootstrap is due to systematic error in the estimation of distances.

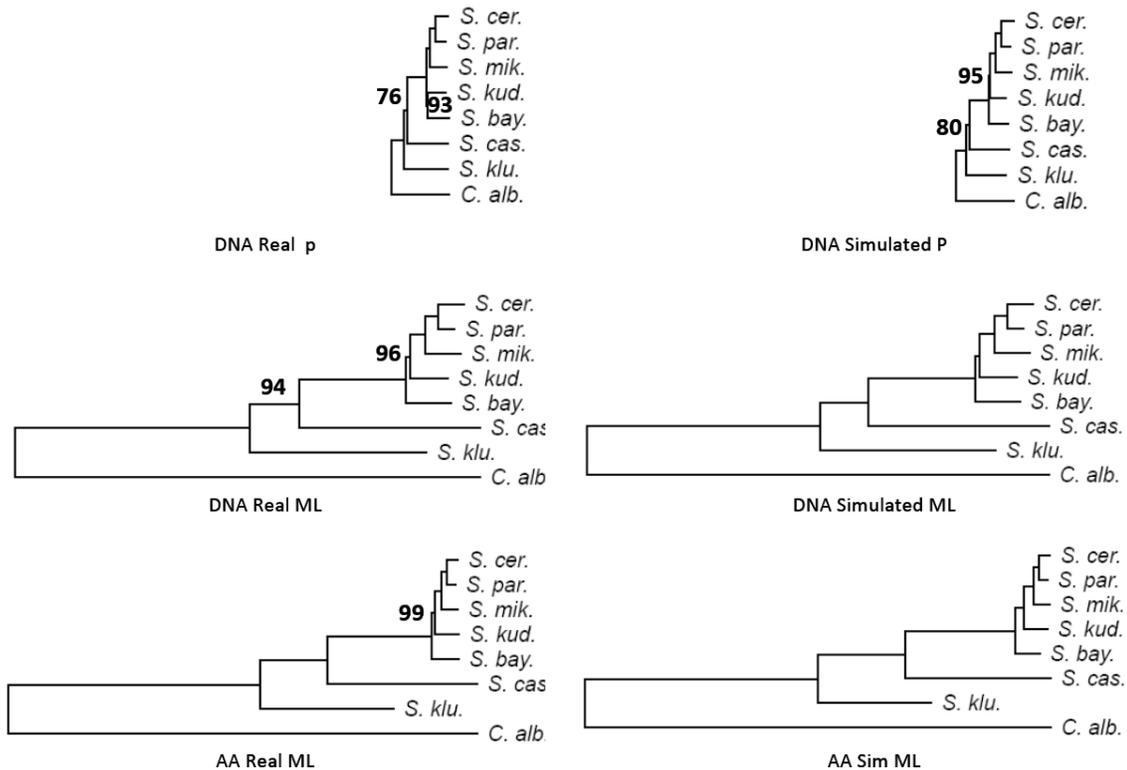

Figure 9. Consensus trees of resampled residual yeast data sets using fWLS for real and simulated distances on DNA and amino acid based distance matrices. For the original data, the optimal tree was found according to the g%SD criterion, and at this *P*-value, the residuals were calculated, standardized and reinflated. Next, 1000 replicate data sets were created by adding these residuals to the expected (optimal tree) distances. For each replicate, the optimal p-value and tree were again found (and the tree with the smallest g%SD kept). Only edges with less than 100% support have values shown. These trees are drawn approximately to scale to emphasize the quite extreme "accordion effect" mentioned in section 3.1 occurring with commonly used distances on this data.

**3.7 Properties of the method when the only errors are stochastic**

It would be very desirable that RR gives similar values to the bootstrap when the only sources of error are sequence length related. That this is the case be expected based on the fact that under the model, sequence resampling, with increasing sequence length, results in multivariate normal errors centered on the generating tree. Thus, if we sample directly from this distribution (for example, as done by Kishino and Hasegawa 1989), then we have a reasonable approximation to bootstrap results. This is indeed the alternative to RELL proposed by these authors. As long as the function $d_{obs}^P$ can make a reasonable approximation to the sampling related errors, and the tree building method can accommodate for ignoring the correlations of errors, then RR should approximate the bootstrap when the model holds. This can be tested using simulations.

An example of comparing RR results with the bootstrap under the model is shown in table 1. The model is the ML DNA model of the yeast data. Using SeqGen 1000 data sets, $X_i$, of 4000 sites were randomly evolved. Each data set was analyzed two ways. In the first, bootstrap approach, the data $X_i$ are immediately resampled once to produce set $Y_i$. This is then analyzed to optimize g%SD, including tree search and finding the optimal value of $P$. The tree with the lowest g%SD is saved and the next replicate is analyzed. This produces a sample of 1000 bootstrap trees. The second approach takes $X_i$ and finds the minimum g%SD tree. At the optimal $P$ value a single RR data set $Z_i$ is produced. For speed, this is analyzed using the same $P$ value and the best tree saved (for accuracy it should be preferable to reoptimize $P$ on the sample). These two sets of 1000 trees are compared in table 1 to show the mean support for each edge in the tree via the two methods. Each set of trees is behaving as a multinomial sample. Clearly the two vectors are very similar, giving a clear indication that in this instance at least the two methods give very similar average results under the model. If anything, in this example, the residual resampling tends to give a bit more support to the best-supported clades than does the sequence bootstrap (this may not be wrong, since the "non-parametric" bootstrap gives less support than the parametric bootstrap in these situations, and the latter appears to be more strictly correct).

Table 1. The frequency of internal partitions (or clades in the rooted tree) in simulated yeast DNA data of length 4000 from 1000 replicates according to Residual Resampling (RR) and the sequence bootstrap when using fWLS. Sequences were simulated according to the ML DNA model and the ML DNA distance used for reconstruction.

| Smaller set in partition | Bootstrap | RR |
|---|---|---|
| *S. cer, S. par* | 100% | 100% |
| *S. cer, S. par, S. mik,* | 100% | 100% |
| *S. cas, S. klu, C. alb* | 100% | 100% |
| *S. cer, S. par, S. mik, S. kud* | 92.4% | 97.2% |
| *S. klu, C. alb* | 90% | 86.4% |
| *S. cas, C. alb* | 5.5% | 9.1% |
| *S. kud, S. bay* | 5.0% | 1.2% |
| *S. Cas, S. klu* | 4.5% | 4.5% |
| *S. cer, S. par, S. mik, S. bay* | 2.6% | 1.6% |

**3.8 Residual Resampling with BME**

Balanced Minimum Evolution (BME) is a criterion that assesses the fit of a tree by calculating the sum of the observed distances across diagnostic paths in the tree being evaluated. It offers some very attractive computational features, for example, those exploited by fastBME, which synchronizes tree search and retrieval of information from the $t^2$ matrix of distances. The other highly desirable feature is that BME seems to be a robust tree inference criterion (Gascuel and Steel 2006). This robustness has been described as BME approximating the weighting of Fitch-Margoliash (FM) Least Squares, or $P = 2$. In turn, it is recognized that the FM+ method (that is, FM limited to non-negative edge lengths) works well in simulations (Kuhner and Felsenstein 1994) and may do so because it approximates Generalized Least Squares (e.g., Gascuel 2000). This last method is known to have desirable statistical properties, such as being a most powerful estimator under asymptotic conditions (e.g., long sequences).

The relationship between BME and FM is sometimes explained by BME being described as a weighted least squares method, where the function being minimized is of the form

$$\sum_{i=1}^{N} \frac{(d_{obs_i} - d_{\exp_i})^2}{2^m} \quad (4)$$

on a binary tree, where $m$ is the number of internal nodes that a distance between two tips crosses (Gascuel and Steel 2006). If a tree is of highly regular shape, e.g. every edge the same length and every lineage splitting evenly, then $2^m$ is a modestly good approximation of $d^2$ on that tree. In the earlier simulations looking at Q-Q plots, NJ trees (being used here as an approximation to the optimal BME tree) showed a poor performance compared to fWLS. In figure 10 we explore why this is the case. Despite $P$ values covering a considerable range, from 1.8 to 3.7, the relationship of s.d. to distance was broadly similar. However, because the optimal tree for the yeast data is both comb-like in shape and has highly dissimilar edge lengths, the BME weight shows a poor colinearity with fWLS weights. Despite this, there is a broad similarity, with the BME weights following a curve of $P \sim 1$.

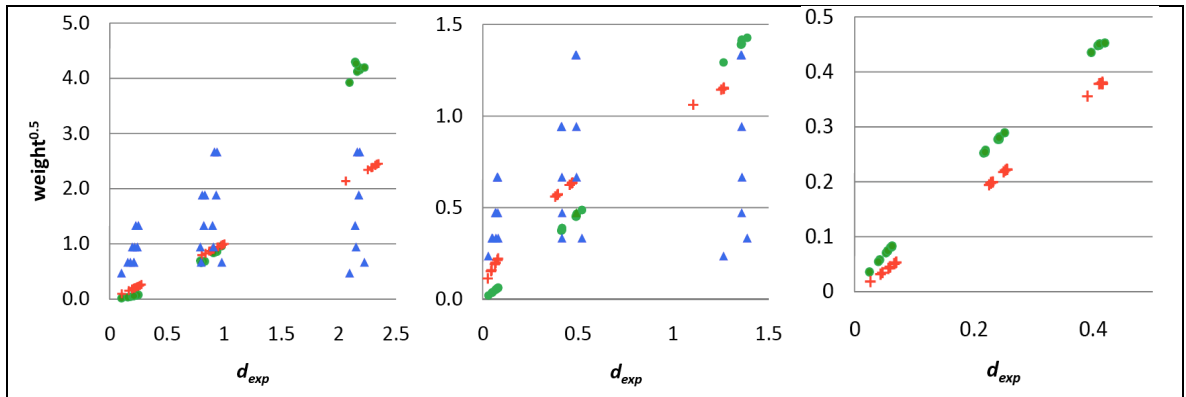

Figure 10. Relationship of weights$^{0.5}$ to path length ($d_{\exp}$) through the tree of yeast data. The green circles are $d_{obs}^P$ for the real data, the red crosses are $d_{obs}^P$ for data simulated under the sequence ML parameters, while blue triangles are $(2^x)^{0.5}$, where $x$ is the number of internal nodes on the tree crossed by a distance, or the BME weight on a binary tree. (a) ML DNA distances on the yeast data $P$ real data = 3.7, simulated data 2.1. (b) ML protein distances on the yeast data $P$ = 2.2 and 1.2, and (c) p protein distances $P$ = 1.8 and 2.2. Note, the BME weights are constant, since the tree is constant, so are not shown on the last figure for this reason.

Figure 11 shows the support residual resampling gives to different edges of the optimal BME (here NJ) tree of the yeast data. Here, the residuals are standardized by dividing by the square root of the BME weight, sampled with replacement, and then redivided by multiplying by the square root of the BME weight on the newly assigned path. Figure 11 may be compared with figure 9 and table 2 with table 1. Again, residual resampling seems to be making a fairly good approximation to the bootstrap values when the model is correct. This is encouraging, as residual resampling combined with fastBME offers the possibility of a fast robust general tree building method, with edge support that reflects the inherent errors in the data as revealed by non-additivity.

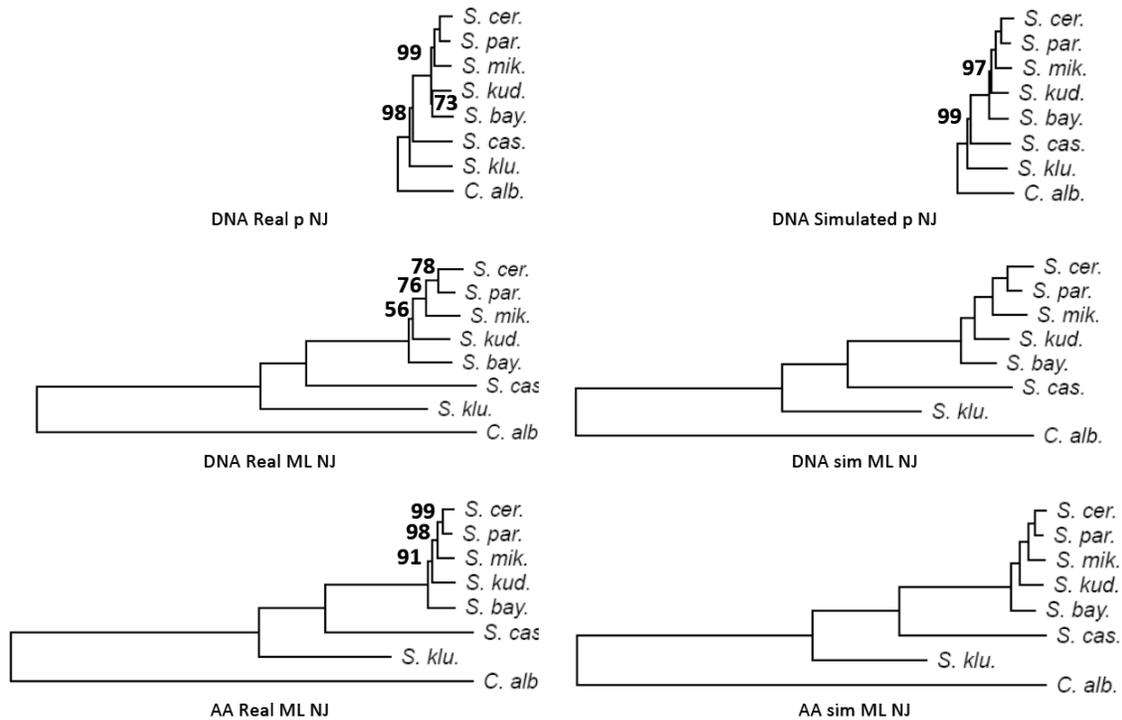

Figure 11. Consensus tree of resampled residual data sets for BME (NJ) for real and simulated data on yeast DNA and amino acid based distance matrices. NJ was used to select the tree on the original data, residuals were then calculated, and standardized as described in section 3.8. Next, 1000 replicate data sets of the expected distances plus residuals were created, and the NJ tree for each kept. Only edges with less than 100% support have values shown. This figure may be compared with figure 9.

Table 2. Average frequency of internal partitions (or clades in the rooted tree) in simulated yeast DNA data of length 4000 from 1000 replicates according to Residual Resampling (RR) and the sequence bootstrap when using NJ. The tree was reconstructed using ML distances.

| Smaller set in partition | Bootstrap | RR |
| --- | --- | --- |
| *S. cas, S. klu, C. alb* | 100% | 100% |
| *S. cer, S. par* | 99.9% | 97.7% |
| *S. cer, S. par, S. mik* | 99.0% | 99.7% |
| *S. klu, C. alb* | 91.0% | 92.9% |
| *S. cer, S. par, S. mik, S. kud* | 88.4% | 94.2% |
| *S. cer, S. par, S. mik, S. bay* | 5.9% | 2.8% |
| *S. kud, S. bay* | 5.7% | 2.9% |
| *S. Cas, S. klu* | 4.5% | 1.1% |
| *S. cas, C. alb* | 4.5% | 6.0% |
| *S. cer, S. par, S. kud, S. bay* | 0.4% | 0.4% |
| *S. cer, S. par, S. kud* | 0.3% | 0.2% |
| *S. cer, S. par, S. kud* | 0.2% | 0.3% |
| *S. mik, S. kud* | 0.1% | 1.2% |
| *S. mik, S. kud, S. bay* | 0.1% | 0.5% |
| *S. par, S. mik* | 0.1% | 0.0% |
| *S. par, S. mik, S. kud, S. bay* | 0.1% | 0.0% |
| *S. cer, S. mik, S. kud, S. bay* | 0.0% | 0.3% |
| *S. cer, S. mik* | 0.0% | 0.1% |

## 4 Discussion

In this article we have explored direct measures of additivity that are equivalent to using likelihood within a distance data set or when $P = 2$ when using flexi Weighted Least Squares (fWLS). They also coincide with a root mean square measure of additivity (percent standard deviation, equation 1) when $P = 2$ and are identical but for the use of a geometric rather than an arithmetic mean. Their use has given indications of which distance transformations are really working best for the purpose of estimating the tree and edge lengths with least total error (stochastic plus systematic error). On the Rokas et al. (2003) yeast genomic data set, they seem to clearly indicate that amino acid distances seem to fit markedly better than nucleotide based distances. Analyses of the residuals indicate that the fit of data to tree with fWLS can appear reasonable, although factors such as the non-independence of residuals (due to the imposed tree structure with distances based on sequences) can result in tests of the normality of residuals, such as Shapiro Wilk, rejecting far to often (and getting much worse with more tips on the tree, results not shown).

Resampling of residuals (RR), where standardization is achieved by dividing by the square root of $d_{obs}$ or ($d_{exp}$) to the power $P$, appears to be a viable alternative to the bootstrap. It has the desirable properties that differences in replicates do not automatically go to zero with increasing sequence length. The present implementation makes the assumption that residuals are uncorrelated. This is indeed an assumption of all the most used distance based tree-building methods, both least squares and BME. It has been argued that, with $P$ close to 2 in value, fWLS will closely mimic the desirable properties of generalized least squares (Gascuel 2000). One obvious way to adjust the resampling of residuals to reflect tree structure is to infer the form of the covariance matrix **V**, based on shared path lengths. Since a vector **y** of correlated heteroscedastic residuals can be created from i.i.d. $N(0,1)$ residuals, **x**, by the transformation **y** = **xV**$^{0.5}$, then **x** can be estimated as **yV**$^{-0.5}$ (ignoring the effect of any boundaries in the parameter space). Another simple addition to RR, is to randomly reverse the sign on the resampled residuals. This improves the accuracy of the resampling if the true distribution of residuals is indeed symmetric. However, given strong convergences (e.g., due to long edges attract) and/or divergences (e.g., long edges repel) affecting certain paths, this may not always be the case. Its use would seem to hold most promise when the tree is relatively small.

Application of residual resampling methods to real data sets highlighted uncertainty in the parts of the tree that do indeed change when better fitting amino acid based evolutionary distances are used. Another way of treating problematic data is recoding DNA sequences into purines (A and G equal R) and pyrimidines (C and T equal Y), so-called RY recoding (e.g., Swofford et al. 1996). By itself, it does not necessarily resolve the hardest problems (e.g. mammalian mtDNA with sparse sampling of species, Waddell et al. 1999), but on this data it has been suggested to reduce systematic error (Phillips, Delsuc and Penny 2004). Applying our methods, we find that an RY p-distance indeed fits much better than four-state DNA distances, with a g%SD of only 0.7547 at an optimal $P$ value of 3.3. Switching to ML RY distances (with a gamma distribution of site rates, shape parameter equals 0.2422), the best tree remains that obtained using protein based distances, but the g%SD decreases further to 0.4832 at $P = 3.1$. Thus there is evidence that the total error about RY distances is even smaller than that obtained when using amino acid distances. In traditional bootstrap analyses, amino acid distances offer the advantage of being a more natural unit for resampling, due to the high correlations in the evolution of the three bases of each codon. This can be accommodated for by using a block bootstrap, where codons (triplets of nucleotides) rather than nucleotides are resampled. However, in the present analysis this is largely irrelevant since residual resampling is directly measuring the total error, and does not need to directly address the difficult question of what sized block to resample. Interestingly, the accordion effect becomes even more pronounced with RY coding than with protein-based distances, further highlighting the great difficulty that accurately estimating edge lengths on phylogenetic trees poses for divergence time estimation (e.g., Kitazoe et al. 2007).

The additivity measures used in this article, together with fWLS offer the potential to compare not just different distance data sets, but also different types of data visualization model (Waddell, Kishino and Ota 2007). The two dominant types of model in data mining are hierarchical clustering (tree building) and multi-dimensional scaling (MDS of which PCA is a popular heuristic approximation). Another type of model worth considering fitting with fWLS and equations 1, 2 or 3 are planar edge weighted graphs, or splits graphs. These are a generalization of trees that appear in applications such as Splits Trees (Huson and Bryant 2006). While there are fast algorithms for calculating the fit of weighted least squares on trees (e.g., Bryant and Waddell 1998), similar methods have yet to be published for Split Decomposition, but the latest versions of the program do use them (David Bryant, pers. comm.). Not only would this allow splits diagrams to be compared directly with trees and MDS, but a likelihood-based criterion like equation 2 or 3 would in turn allow testing of how many edges to include in the splits diagram. When combined with a tree metric, all these visualization methods, can in turn be used to visualize sets of trees in a higher-level meta-analysis (e.g., Waddell and Kishino 2001).

Maximizing additivity may hold the key to selecting the most appropriate distance measure (model) to use when estimating a tree. This is because it monitors increasing stochastic error in distance estimation just as much as (hopefully) reduced systematic error for more general distances. It also offers hope of avoiding some of the risks associated with long edge repulsion, due, for example to non-stationarity of character frequencies (Waddell 1995). In this article, this appears as a potential reason why pairwise GTR distances may perform worse than ML distances with real data.

Some of the concepts explored here can also be applied to other model-based methods of phylogenetic analysis. Examples are minimum $G^2$ (maximum likelihood), minimum $X^2$ methods, or weighted least squares Hadamard methods (Waddell 1995). The idea here is to resample the residuals seen in either **s** (sequence) space, or **γ** (tree) space. When using weighted least squares with $\hat{\gamma}$, the tree fits perfectly to the set of edges contained in it, but all the other edge length residuals, i.e. $\hat{\gamma}_i$/s.d., are not fitted (here, s.d. may be estimated by the raw data, or with more sensitivity under the model by projecting to **s** space then inferring **V** of gamma, Waddell et al. 1994). The gamma vector can be recentered to a tree by setting all non-tree entries to zero, then randomly reassigning the residuals. For the minimum $X^2$ method, the resampling of standardized residuals might be shifted to the sequence space, with or without recentering. With minimum $X^2$ and likelihood, the sum of all residuals in theory, contribute to inflate the overall model fit statistic (e.g. the likelihood ratio between the tree model and the i.i.d. multinomial model). If bootstrap resampling is then performed, the researcher may want to aim to replicate this discrepancy by reducing the number of resampled site patterns (that is, adjust the effective sequence length). This is loosely related to the concept in quasi-likelihood, that is, inflate the model variance to match the level of mismatch seen for the overall model. However, there are suggestions that the likelihood ratio statistic has weak power in sequence analyses to pick up relevant deviations from the model (Waddell, Ota and Penny 2009). In such cases, various types of marginal likelihood or $X^2$ tests may better detect the deviations expected to affect tree estimation with real data. If these are deemed relevant to the accuracy of the phylogenetic model (as things like base composition stationarity indeed seem to be) then these statistics may stand in as the target for "variance matching."

Since the intermediate vectors of the Hadamard, that is **r** and **ρ** are, respectively, linear transforms of observed and expected "generalized" distances (e.g., Waddell et al. 1994), they too might be targets for residual resampling (specifically targeting **ρ**). After a tree is specified and fitted to yield $\hat{\gamma}$(T), it may be back transformed to give the expected vector, $\hat{\rho}$(T). Comparing this with the observed, $\hat{\rho}$, a full set of residuals are then ready to be resampled. This is another way of allowing the data, rather than just the model, to suggest the scale and distribution of error. Unlike pairwise distances, but like sequence-pattern likelihood, the generalized distances of the Hadamard, allow detection of monotonic error even when a molecular clock holds (Waddell 1995). This desirable property does not apply to the distance Hadamard (Hendy and Penny 1993), which behaves more like a standard distance method (Waddell 1995).

Another direction worth exploring is replacing $d^P$ as being proportional to variance, with another function of $d$ chosen to more closely match theoretically calculated variances to predicted distances. This may suffer from two weaknesses; it may not be amenable to the fastest forms of WLS, which can reduce the optimization of edge lengths on a single tree from $t^3$ to $t^2$ after the first tree, by reusing many of the calculations (as implemented in PAUP, David Bryant pers. comm.). This acts to reduce the overall complexity when the tree search is at least order $t$ itself, as it almost always is. Secondly, since most real data contains complex combinations of stochastic and systematic errors that are fitted together, the fitting using $d^P$ may often be equally good (or bad) in practice.

Distance and ML methods can also be combined in other ways and need not be thought of as simply either or alternatives. In many cases, sequence based ML seems to have more robustness than distance based methods (e.g., Waddell 1995). However, the very fine-grained optimization it performs may erase some of the telltale signs of misfit. One way to test the assumptions of an ML sequence model is to compare how model based distances fit to the real data, using just the types of simulation performed in figure 2 for example. Distances by virtue of trying to directly infer long paths through the tree, may often be more vulnerable than ML (which has internal nodes as reference points), but that very property allows them to serve as a "canary in the coal mine" for those intrepid molecular data miners that trust implicitly whatever answer ML returns.

Finally, it worth bearing in mind that residual resampling is not a foolproof method and there is the possibility, however slight, that even an editor of Nature or Science Magazine might misinterpret the results. It does however seem to be an improvement that should give the reader a better sense of the robustness of the tree topology. In cases where systematic error is expected, then there needs to still be an onus on the investigator(s) to attempt to identify its source(s) and mitigate its effect. Despite the popularity of software that may seem to demand minimal choices of the operator, phylogenetic analysis is not quite a push button process.

"… and with the ignorance he's got that makes him one of the most powerful men that have ever lived." A quotation from Kryten: Red Dwarf VIII-Cassandra

*Software to apply residual resampling, working with the PHYLIP package is available from the authors.*

## Acknowledgements

This work was supported by NIH grant 5R01LM008626 to PJW. Thanks to Joe Felsenstein, Olivier Gascuel, Hiro Kishino, Mike Steel, and Dave Swofford for helpful discussions. Special thanks to Dave Swofford for use of a test version of PAUP that offers any real value for parameter *P* when using its least squares functions, and Alex Pothen for allowing AA to share time on this project while working on another project.

## Author contributions

PJW originated the research, developed methods, gathered data, interpreted analyses, prepared figures and wrote the manuscript. AA implemented methods in C and PERL, ran analyses, prepared figures, interpreted analyses and commented on the manuscript.